\documentclass[abbrv,aps,prb,twocolumn,showpacs,preprintnumbers,amsmath,amssymb,
superscriptaddress,nofootinbib]{revtex4}

\usepackage{amsmath,amssymb,amsfonts}
\usepackage{bm,graphicx,color}

\newcommand{\laco}[1]{$\mathrm{LaCoO_3 }$}

\def\lsim{~\rlap{$<$}{\lower 1.0ex\hbox{$\sim$}}}
\def\gsim{~\rlap{$>$}{\lower 1.0ex\hbox{$\sim$}}}

\begin{document}
\title{Phase diagram of exciton condensate in doped two-band Hubbard model}
\author{Jan Kune\v{s}} 
\affiliation{Institute of Physics, Academy of Sciences of the Czech republic, Cukrovarnick\'a 10,
Praha 6, 162 53, Czech Republic}
\email{kunes@fzu.cz}
\date{\today}

\begin{abstract}
Using the dynamical mean-field approximation we investigate formation
of excitonic condensate in the two-band Hubbard model in the vicinity 
of the spin-state transition. With temperature and band filling as the
control parameters we realize all symmetry allowed spin-triplet excitonic
phases, some exhibiting a ferromagnetic polarization.
While the transitions are first-order at low temperatures,
at elevated temperatures continuous transitions are found that give rise
to a multi-critical point. Rapid but continuous transition between 
ferromagnetic and non-magnetic excitonic phases allows switching 
of uniform magnetization by small changes of chemical potential.
\end{abstract}
\maketitle

\section{Introduction}
Strongly correlated fermions attract interest  
for their strong response to various external perturbations. 
This feature is often connected to existence of competing phases 
which can be switched by small changes of physical parameters. 
One way to materials with rich phase diagrams is to look for 
ordered states characterized by order parameters with complex structure. 
A prototypical example is superfluidity of $^3$He 
where the order parameter is a complex
$3\times3$ matrix, resulting in numerous thermodynamic phases~\cite{vollhardt}.
Common types of order such as charge and spin density waves 
or s-wave superconductivity usually support only a single phase for a
given translational symmetry. Ordered states that allow multiple phases
such as p-wave superconductor or excitonic condensate (EC) are more exotic.

The idea of excitonic instability at the semiconductor-semimetal transition 
was introduced by Mott in 1961~\cite{mott61} and 
the early developments on the topic were summarized 
by Halperin and Rice~\cite{halperin68a,halperin68b}.
More recently the EC concept
was used to study hexaborides~\cite{balents00a,balents00b,barzykin00,veillette01}
and bi-layer systems~\cite{eisenstein04,rademaker13a,rademaker13b}, where the EC is driven 
by inter-atomic Coulomb interaction. A possibility of EC driven by
intra-atomic Coulomb interaction was proposed recently for 
$5d^4$ and $3d^6$ perovskites.~\cite{khaliullin13,kunes14b}

A minimal model of EC consists of two electronic bands with a small gap/overlap
close to half filling and an inter-band Coulomb interaction providing the pairing
glue. The theory of EC is formally similar to the theory of superconductivity. However,
since the EC order parameter is orbital off-diagonal it may possess both the spin-singlet and spin-triplet
components even when being local ($s$-wave). For weak exchange, typical
of inter-atomic interaction, the singlet and triplet excitons are of comparable energy and the
order parameter has four complex components~\cite{halperin68b}, allowing numerous 
thermodynamic phases. In particular, excitonic ferromagnetism may arise from mixing of 
the singlet and triplet pairing states~\cite{volkov75}. Strong ferromagnetic exchange,
materialized in intra-atomic Hund's coupling, suppresses the singlet 
pairing state.
While the number of the permissible phases is thus reduced it remains larger than one.

Starting with the early theoretical studies, the weak-coupling approach leading to a
BCS-like mean-field theory has been used to describe the EC
~\cite{descloizeaux65,balents00a,veillette01,zocher11}. The strong-coupling approach was
taken by Balents~\cite{balents00b}. A special instance of the strong-coupling EC theory
is obtained for bi-layer Heisenberg model treated in bond-basis.~\cite{sachdev90,sommer01,giamarchi08}
Heavier numerical methods that do not rely on expansions in the interaction strength
have only recently been employed. Rademaker~{\it et al.}~\cite{rademaker13a,rademaker13b} used 
quantum Monte-Carlo and Kaneko~{\it et al.}~\cite{kaneko12} used variational cluster
approximation to investigate the EC in two-band Hubbard model without the exchange interaction.
Kune\v{s} and Augustinsk\'y performed a dynamical mean-field calculations on the two-band Hubbard
model with strong Hund's exchange and found an excitonic instability close
to the spin-state crossover~\cite{kunes14a}. Subsequently the calculations were 
extended to study the physics of the ordered phase below the excitonic $T_c$.~\cite{kunes14b}
The role of Hund's exchange in selecting the singlet and triplet EC order
was addressed in variational cluster study by Kaneko and Ohta~\cite{kaneko14}.


The work reported in this Article is a continuation of our effort to
map the instabilities occurring in systems close to the spin-state crossover
as described by the two-band Hubbard model. Close to the
half-filling the physics of the model is governed
by the competition between the crystal field and Hund's exchange.~\cite{werner07,suzuki09} 
In the vicinity of the high-spin--low-spin crossover the degeneracy of the 
corresponding atomic states leads to long-range ordering. 
For strongly asymmetric bands, the system behaves as the classical Blume-Emery-Griffiths
model~\cite{beg} with high-spin--low-spin order when on bipartite lattice.~\cite{kunes11}
For less asymmetric bands, the quantum effects gain on importance and the system
becomes unstable towards excitonic condensation.~\cite{kunes14a}
Due to its complex nature the excitonic pairing allows several distinct thermodynamic phases.
In this work, we map out the EC phase diagram as a function of
temperature and band filling. Going beyond the previously studied doping levels
we are able to observe all the symmetry allowed phases. In particular, in addition
to the spin-density-wave phase observed close to half-filling~\cite{kunes14b,kaneko14}
we observe two ferromagnetic phases.

\section{\label{sec:comp}Computational procedure}
We consider the two-band Hubbard model with nearest-neighbor (nn) hopping on a bipartite (square) lattice
with the kinetic
$H_{\text{t}}$ and the interaction $H_{\text{int}}=H^{\text{dd}}_{\text{int}}+H'_{\text{int}}$ terms given by
\begin{equation}
\label{eq:hubbard}
\begin{split}
&H_{\text{t}}=\frac{\Delta}{2}\sum_{i,\sigma}\bigl(n^a_{i\sigma}-n^b_{i\sigma}\bigr)+
  \sum_{i,j,\sigma}\bigl(t_{a} a_{i\sigma}^{\dagger}a^{\phantom\dagger}_{j\sigma}+
t_{b} b_{i\sigma}^{\dagger}b^{\phantom\dagger}_{j\sigma}\bigr) \\
&\qquad+\sum_{\langle ij\rangle,\sigma}\bigl(V_1a_{i\sigma}^{\dagger}b^{\phantom\dagger}_{j\sigma}+
V_2b_{i\sigma}^{\dagger}a^{\phantom\dagger}_{j\sigma}+c.c.\bigr) \\
&H^{\text{dd}}_{\text{int}}=U\sum_i \bigl(n^a_{i\uparrow}n^a_{i\downarrow}+n^b_{i\uparrow}n^b_{i\downarrow}\bigr)+
  (U-2J)\sum_{i,\sigma} n^a_{i\sigma}n^b_{i-\sigma}\\
&\qquad+(U-3J)\sum_{i\sigma} n^a_{i\sigma}n^b_{i\sigma} \\
&H'_{\text{int}}= J \sum_{i\sigma} a_{i\sigma}^{\dagger}b_{i-\sigma}^{\dagger}
a_{i-\sigma}^{\phantom\dagger}b_{i\sigma}^{\phantom\dagger} 
+J'\sum_{i} \bigl(a_{i\uparrow}^{\dagger}a_{i\downarrow}^{\dagger}b_{i\downarrow}^{\phantom\dagger}
b_{i\uparrow}^{\phantom\dagger}+c.c.\bigr).
\end{split}
\end{equation}
Here $a_{i\sigma}^{\dagger}$, $b_{i\sigma}^{\dagger}$
are the creation operators of fermions with spin $\sigma=\uparrow,\downarrow$
and $n^c_{i\sigma}=c_{i\sigma}^{\dagger}c^{\phantom\dagger}_{i\sigma}$.
The parameters of the Hamiltonian are the same as in Refs.~\onlinecite{kunes14a,kunes14b}:
$U$=4, $J$=1, $\Delta=3.40$, $V_{1,2}=0$, $t_a=0.4118$, and $t_b=-0.1882$. We assume
eV to be the unit of energy and express temperature in Kelvin.
The choice $t_at_b<0$ ensures
a ferro EC order \cite{kunes14a} allowing us to work with a one-site unit cell.

The numerical calculations were performed in the dynamical mean-field approximation~\cite{dmft,metzner89}
with the density-density interaction $H^{\text{dd}}_{\text{int}}$ only.
The effect of adding $H'_{\text{int}}$ is considered in Section~\ref{sec:su2}. 
The auxiliary impurity problem is solved using the hybridization-expansion continuous-time quantum
Monte Carlo (CT-HYB)~\cite{werner06,ctqmc} in the so called segment implementation,
modified to allow off-diagonal hybridization. The spectral functions were obtained
with maximum entropy method.~\cite{gubernatis91}

The convergence of the DMFT loop for parameters close to the phase boundaries required several 
thousands of iterations (typical value far from the boundaries was 20-40 iterations)
and was usually checked by starting the iterative procedure in both phases.
\begin{figure}
\includegraphics[width=0.7\columnwidth,angle=270,clip]{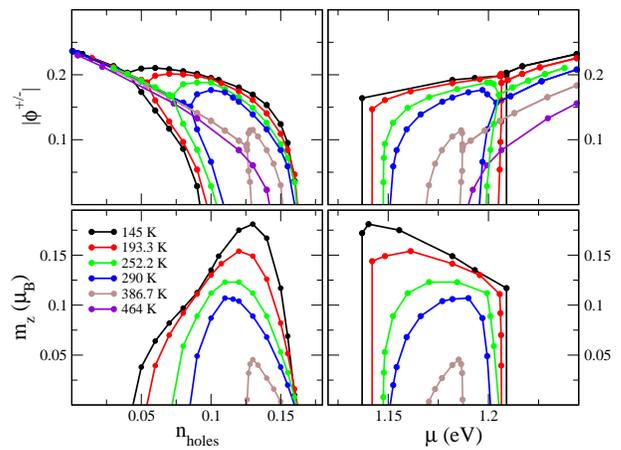}
\caption{ \label{fig:T-cut} (color online)
Dependence of the order parameters $|\phi^+|$ and $|\phi^-|$ (top row) and the
net magnetization (bottom row) on the the doping $n_h$ (left) and chemical potential
$\mu$ (right) for several temperatures. We point out that $n_h$-dependences for $(n_h,T)$
falling into the phase separation regime, which correspond to an unphysical (unstable) phase,
are not distinguished in the plot.
}
\end{figure}
\begin{figure}
\includegraphics[width=0.45\columnwidth,angle=270,clip]{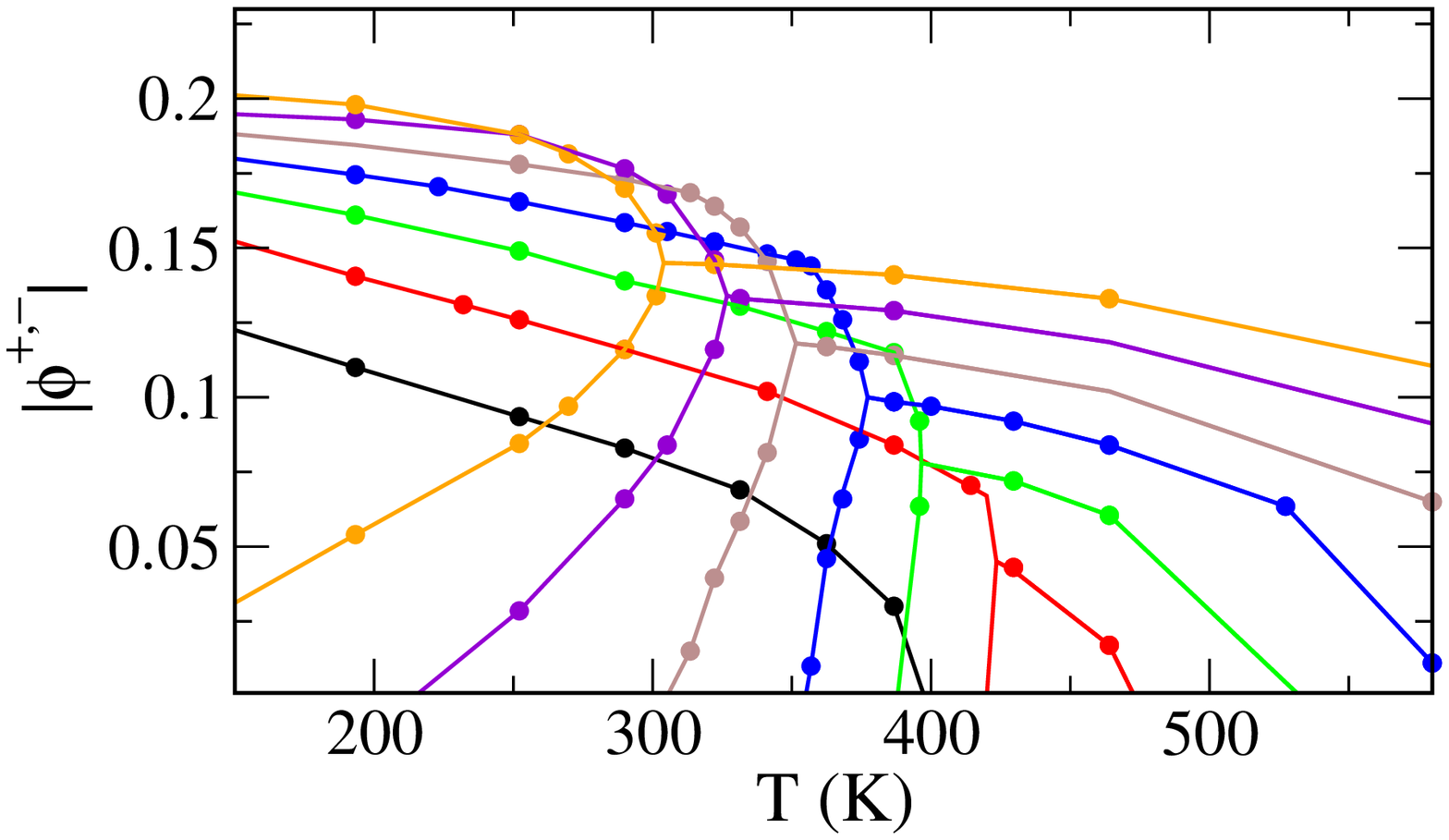}
\includegraphics[width=0.45\columnwidth,angle=270,clip]{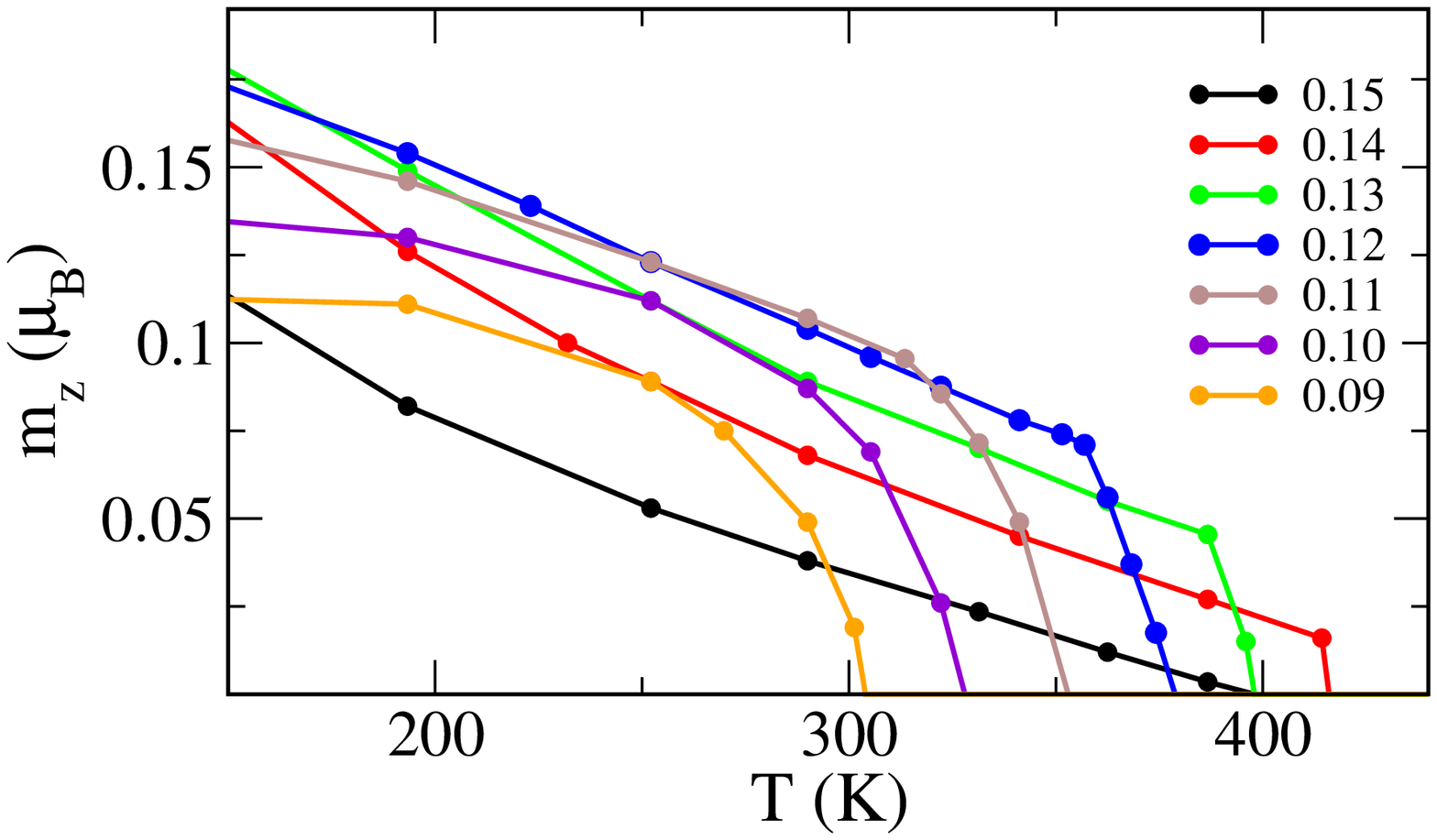}
\caption{ \label{fig:n-cut} (color online)
The temperature dependence of the order parameters $|\phi^+|$ and $|\phi^-|$ (top) and the
net magnetization (bottom) at fixed doping. We point out that results for $(n_h,T)$
falling into the phase separation regime, which correspond to an unphysical (unstable) phase,
are not distinguished in the plot.
}
\end{figure}
\section{\label{sec:num}Numerical results}
In this section we present the DMFT results obtained for
the Hamiltonian $H_t+H^{\text{dd}}_{\text{int}}$. 
The variable parameters are the temperature $T$ and the hole doping $n_h$.
The excitonic condensate can be characterized by a complex 2-component local
order parameter $\mathbf{\phi}=(\phi^x,\phi^y)$ 
\begin{equation}
	\phi^{\gamma}=\sum_{\alpha\beta}\sigma_{\alpha\beta}^{\gamma}\langle a^{i\dagger}_{\alpha}b^{i\phantom\dagger}_{\beta}\rangle,
\end{equation}
where $\sigma^{\gamma}$ ($\gamma=x,y$) are the Pauli matrices.
Throughout the paper we will use the representation 
$(\phi^+,\phi^-)=(\phi^x+i\phi^y,\phi^x-i\phi^y)/2$, which is more practical for 
our purposes. Due to the symmetry of Hamiltonian (\ref{eq:hubbard})
a variation of the phase of either $\phi^+$ or $\phi^-$ generates degenerate states.
Different thermodynamic phases are, therefore, distinguished by the amplitudes $|\phi^+|$ and $|\phi^-|$.

\subsection{Order parameter and magnetization}
\begin{figure}
\includegraphics[width=0.7\columnwidth,angle=270,clip]{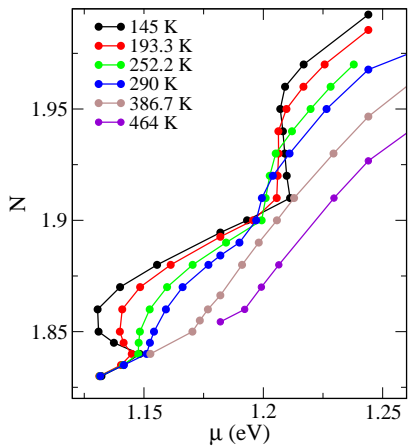}
\caption{ \label{fig:nmu} (color online)
	The $n(\mu)$ dependence of the charge density on the chemical potential
	for various temperatures. The line are guides to the eye connecting the 
	points linearly.
 }
\end{figure}

In Figs.~\ref{fig:T-cut}a, \ref{fig:n-cut}a we show the evolution of $|\phi^+|$ and $|\phi^-|$ along
several constant--$n_h$ and constant--$T$ scans. Four distinct phases can be identified:
i) the normal (N) phase  $|\phi^+|=|\phi^-|=0$, ii) the linear (L) excitonic phase $|\phi^+|=|\phi^-|\neq0$, 
iii) the circular (C) excitonic phase $|\phi^+|=0$, $|\phi^-|\neq0$ and iv) the elliptic (E) excitonic phase 
$0\neq|\phi^+|\neq|\phi^-|\neq0$.  
For each $(n_h,T)$ pair we found a unique solution. This is, however, not the case for fixed 
chemical potential $\mu$ at low temperatures. In Fig.~\ref{fig:nmu} we show the electron density per 
atom $N$ as a function of $\mu$. The plot reveals that some solutions 
obtained at constant $n_h$ are unstable (negative compressibility). Using the Maxwell construction
we can identify the charge separation regions and the first-order transition lines.
The order parameter as a function of $\mu$ is shown in Fig.~\ref{fig:n-cut}b.

The C and E phases exhibit a finite uniform magnetization $\langle m_z \rangle$ shown 
in Figs.~\ref{fig:T-cut}c,d and Fig.~\ref{fig:n-cut}b. The  ordered magnetization exhibits several unusual 
features. While $\phi$ follows the square root $(1-\tfrac{T}{T_c})^{1/2}$ dependence expected of
a mean-field order parameter, $\langle m_z \rangle\sim1-\tfrac{T}{T_c}$ behaves linearly in the vicinity 
of the C/N boundary (black line in Fig.~\ref{fig:n-cut}b). Below the L/E boundary
the $\langle m_z \rangle$ appears to follow the square root dependence (orange line in Fig.~\ref{fig:n-cut}b).
While these observations based on a few data points are somewhat speculative 
we present supporting physical arguments in the discussion. Another feature we want to mention
is the great sensitivity of $\langle m_z \rangle$ to variations of the chemical potential
in the vicinity of the E phase. This property could be used in construction of
devices where magnetization is controlled by a gate voltage.

\subsection{Spectral density}
\begin{figure}
\includegraphics[width=0.7\columnwidth,angle=270,clip]{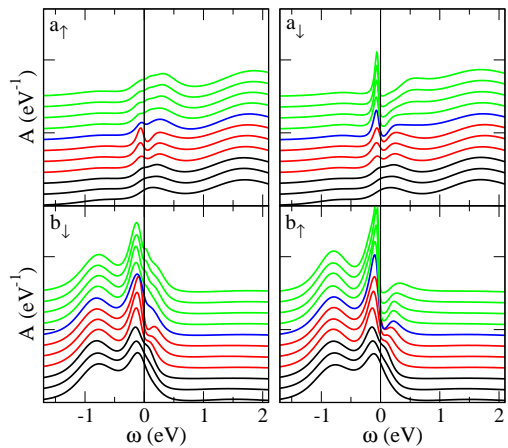}
\caption{ \label{fig:spec} (color online)
The evolution of one-particle spectra with temperature at fixed doping
of 0.12 holes/site. The panels show the diagonal spectral functions
$A_{\alpha\alpha}(\omega)$ ($\alpha=a_{\uparrow}$, $a_{\downarrow}$, $b_{\uparrow}$, $b_{\downarrow}$)
at various temperatures (offset for clarity).
The temperature increases from the top (223.1~K) to the bottom (828.6~K), 
the colors distinguish the different phases: N (black; 828.6~K, 725~K, 594.9~K), L (red;
527.3~K, 464~K, 386.7~K), E (blue; 362.5~K), and
C (green; 341.2~K, 290~K, 252.2~K, 223.1~K).
}
\end{figure}

In Fig.~\ref{fig:spec} we show the evolution of the one-particle spectral density 
as a function of $T$ for fixed doping $n_h=0.12$. All the four phases are traversed as the
temperature is varied. Similar to the behavior of the undoped system, 
reported in Ref.~\onlinecite{kunes14a}, the appearance of the off-diagonal self-energy 
in the EC phase causes a dip in the spectral function, a precursor of a gap.
In the L phase the doping prevents the chemical potential from being inside the gap.
In the C phase, the system approaches a half-metal-like state where the condensed channel
($a_{\downarrow}$ and $b_{\uparrow}$ for  $|\phi^-|\neq0$) 
is gapped and hosts one electron per site~\cite{note1},
while the uncondensed channel ($a_{\uparrow}$ and $b_{\downarrow}$) serves 
as an electron reservoir. Note, nevertheless, that the changes to the $a_{\uparrow}$ and
$b_{\downarrow}$ spectral densities in the C phase go beyond a simple
rigid band shift of the corresponding normal phase spectra.
For non-zero cross-hopping $V_{1,2}$ the condensed and uncondensed channels cannot
be strictly defined and this picture applies only approximately.

\subsection{Phase diagram}
\begin{figure}
\begin{center}
\includegraphics[width=0.7\columnwidth,angle=270,clip]{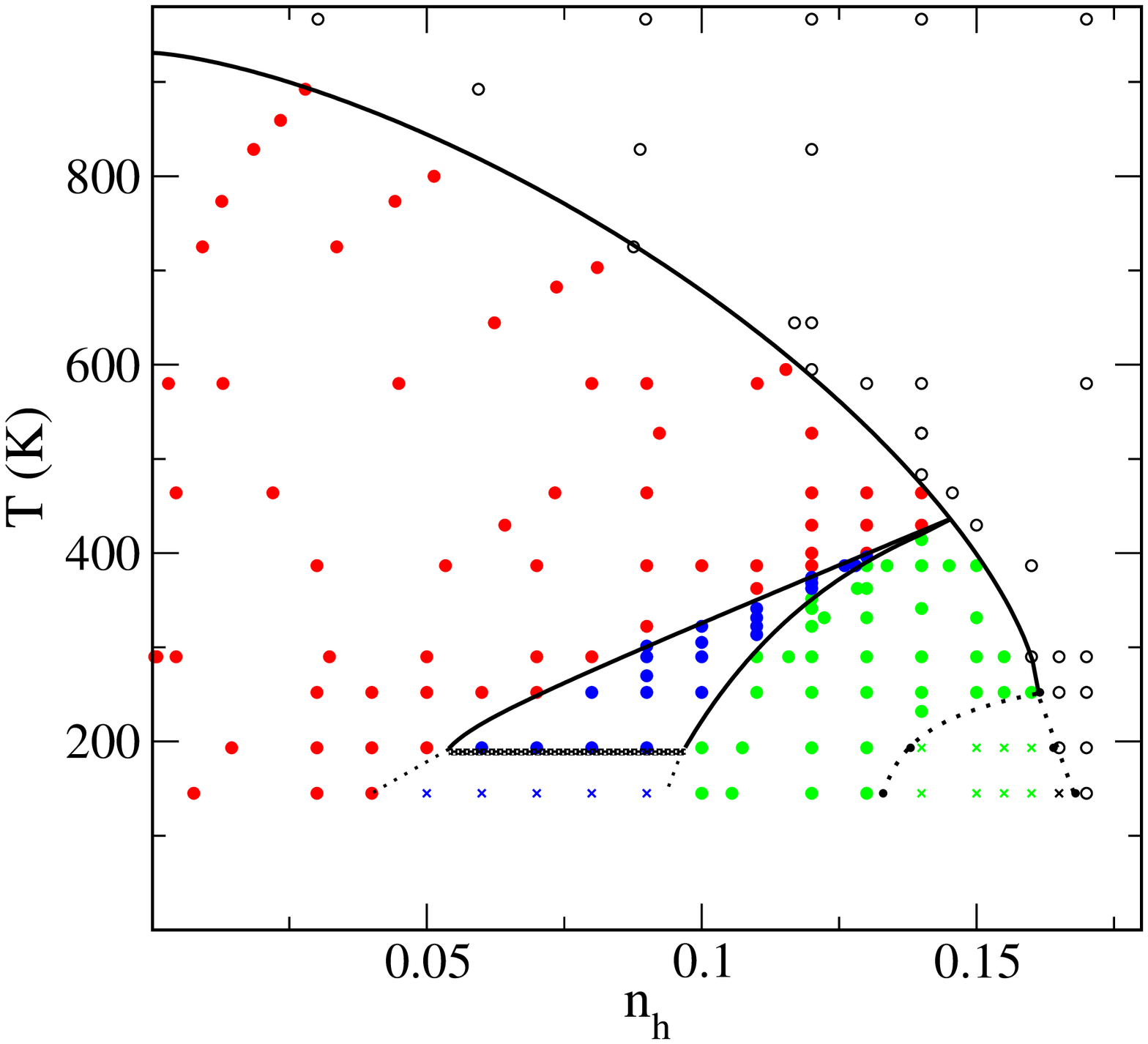}
\includegraphics[width=0.7\columnwidth,angle=270,clip]{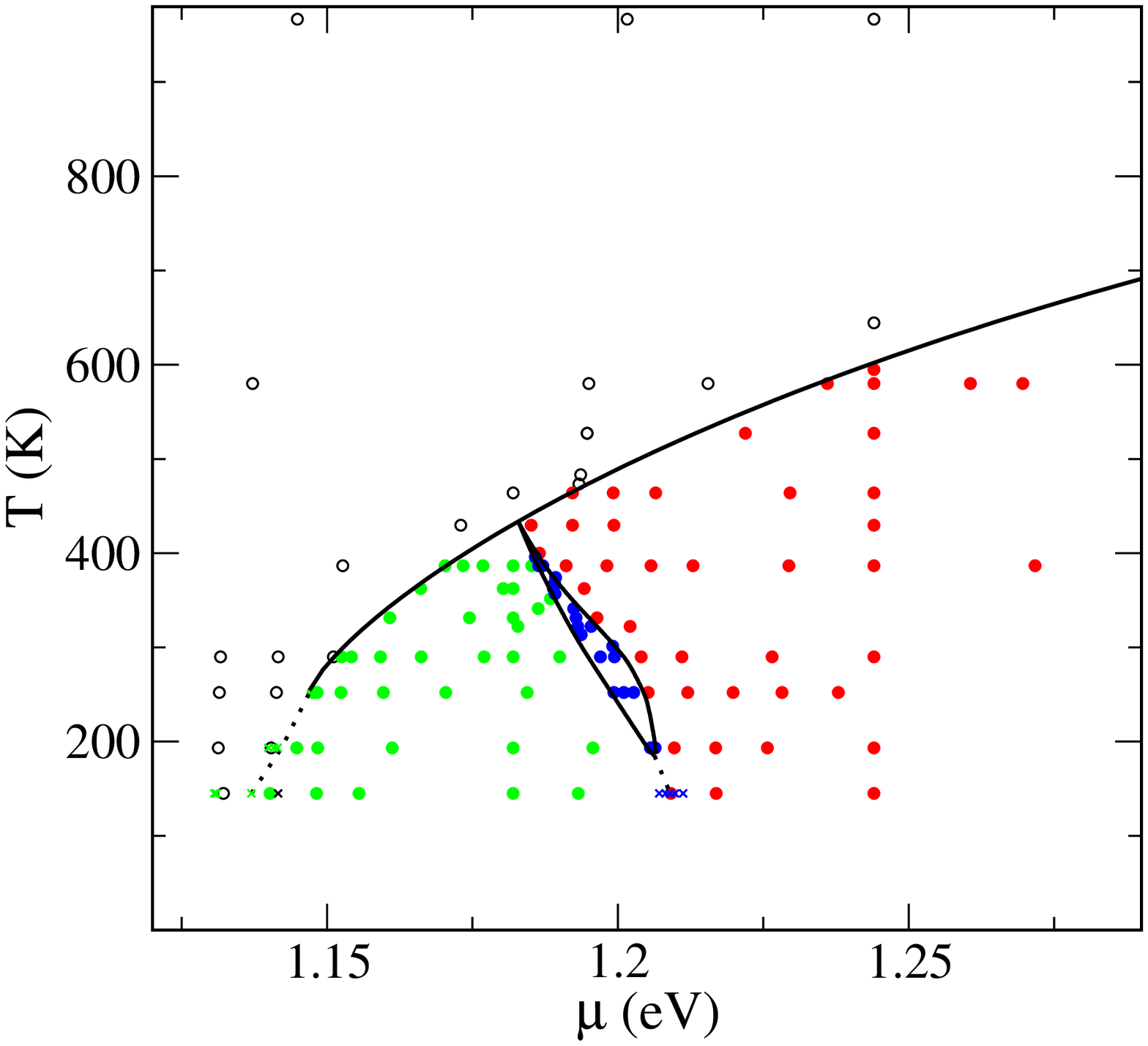}
\end{center}
\caption{ \label{fig:phase_diag} (color online)
Top: the phase diagram in the density-temperature ($n_h-T$) plane. The symbols correspond
to the parameters where actual calculations were performed. The circles mark stable solutions
while the crosses the thermodynamically unstable ones. The colors code the thermodynamic phases:
N (open circles), L (red), E (blue), and C (green) 
The lines mark the estimated phase boundaries corresponding to continuous transitions
(solid) and the phase separation region (dotted).
Bottom: the corresponding phase diagram in the $\mu-T$ plane. The solid lines mark the continuous
transitions, while the dotted lines the first order ones.
}
\end{figure}

In Fig.~\ref{fig:phase_diag} we summarize our main result, 
the phase diagrams in the $n_h-T$ and $\mu-T$ planes.
The undoped system undergoes a continuous transition to the L phase where it stays
down to the lowest studied temperature (145~K). The instability of the normal phase 
shifts to lower $T_c$ with doping and excitonic phase disappears completely above
0.17~holes/site. The continuous N/L transition changes into a continuous 
N/C transition around the doping of 0.145~holes/site. At even higher dopings 
the N/C transition becomes first-order. The E phase is found in a narrow wedge
between the L and C phases. Below approximately 190~K the E phase disappears
and the system evolves between L and C phases through a first-order transition.

All four phases appear to come together at a multi-critical point $(n_h,T)\approx$(0.145,430~K). 
It is not possible for our numerical method to capture details of this region, 
e.g., to exclude a short but finite E/N boundary. Nevertheless, we present 
an analysis based on the Landau theory, 
which supports the phase diagram as drawn in Fig.~\ref{fig:phase_diag}.

\subsection{Landau functional}
The symmetry of our model restricts the form of the Landau functional to
\begin{equation}
\label{eq:landau1}
\begin{split}
        &F(|\phi^+|,|\phi^-|)=\alpha\left(|\phi^+|^2+|\phi^-|^2\right)\\
        &+\beta_0\left(|\phi^+|^2+|\phi^-|^2\right)^2+\beta_1\left(|\phi^+|^2-|\phi^-|^2\right)^2+\ldots,
        \end{split}
\end{equation}
where the higher oder terms contain all powers of $|\phi^+|^2+|\phi^-|^2$ and
even powers of $|\phi^+|^2-|\phi^-|^2$. The terms up to 4th order in $\phi$ 
are sufficient to show that for a finite $\beta_1$
a continuous transition to either C phase (for $\beta_1<0$ and $\beta_0>-\beta_1$)
or L phase (for $\beta_1>0$ and $\beta_0>0$) is obtained.
The L/N and C/N second-order phase boundaries, therefore, meet at a single point ($\beta_1=0$). 
The critical end point of the C/N boundary corresponds to $\beta_0=-\beta_1$.

\begin{figure}
\includegraphics[width=0.4\columnwidth,angle=270,clip]{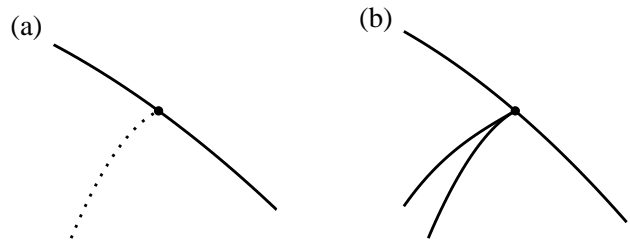}
\caption{ \label{fig:cartoon}
The two possible shapes of the phase boundaries in the vicinity of the multi-critical point.
(a) A first-order transition between C and L phases, (b) two second-order transitions
with E phase in between. The second-order transition lines (or the boundaries of the co-existence
region) approach the multi critical point with the same slope.
}
\end{figure}

The analysis of the phase diagram in the vicinity of the multi-critical point $\beta_1=0$
is more complicated and involves the terms up to order $\phi^8$. We postpone the 
details to the Appendix and here present only the main result. 
The functional form (\ref{eq:landau1}) restricts the behavior in
the vicinity of the multi-critical point to two possibilities shown in Fig.~\ref{fig:cartoon}: i) a first-order transition
between the C and L phases, ii) two continuous transitions with an intermediate E phase.
Moreover, the C/E and L/E boundaries approach the multi-critical point with the
same slope. The terms up to the 8th order decide which of the two scenarios is realized.
In our model the numerical data show a clear preference for the scenario (ii).

\subsection{Symmetry considerations}
\label{sec:sym}
Another way to gain insight into the possible phases of the studied model is to
analyze the transformation properties of the order parameter $\boldsymbol{\phi}$
under the symmetries of the Hamiltonian (\ref{eq:hubbard}). We will also consider 
the qualitative changes to the phase diagram
when external magnetic field or cross-hopping $V_{1,2}$ is added.
The Hamiltonian (\ref{eq:hubbard}) is invariant under a phase change 
in any of the four spin-orbit flavors. 
In addition there is a discrete $\mathbb{Z}_2$ symmetry 
with respect to a $\pi$-spin-rotation (simultaneous
in the $a$ and $b$ orbitals) about an arbitrary axis perpendicular to $z$-axis.
Taken together the symmetry group can be written as
$\Gamma=U_{c_a}(1)\times U_{c_b}(1)\times \{[U_{s_a}(1)
\times U_{s_b}(1)]\rtimes \mathbb{Z}_2\}$,
where the first two factors correspond to conservation of charge in the $a$ and $b$-channels,
the latter two $U(1)$ factors correspond to conservation of the 
$z$-component of spin in the $a$ and $b$-channels
and form a semi-direct product with the 
$\mathbb{Z}_2$ group containing the $\pi$-spin-rotation, e.g., $\{I,\sigma_x\}$.
The order parameter transforms as
\begin{equation}
	\label{eq:transf}
\begin{split}
		R
		(\boldsymbol{\varphi})
	\left(
		\begin{array}{c} \phi^+ \\ \phi^- 
        \end{array}\right)
	&=
	e^{i(\varphi_{c_a}-\varphi_{c_b})}\left(
\begin{array}{c} e^{i(\varphi_{s_a}+\varphi_{s_b})}\phi^+ \\
		         e^{-i(\varphi_{s_a}+\varphi_{s_b})}\phi^-
 \end{array}\right) \\
 \sigma_x\left( \begin{array}{c} \phi^+ \\ \phi^- \end{array}\right)& =
 \left( \begin{array}{c} \phi^- \\ \phi^+ \end{array}\right)
 \end{split}
\end{equation}
under the rotations (phase variation) $R$ and the $\pi$-spin-rotation
about the $x$-axis.

Distinct phases can be found by inspecting the 
operations which leave the order parameter $\boldsymbol{\phi}$ invariant. 
Besides the trivial solution corresponding to the normal phase there are three distinct 
solutions: (i) $0\neq \phi^+ \neq \phi^- \neq 0$, implying $\varphi_{c_a}=\varphi_{c_b}$
and $\varphi_{s_a}=-\varphi_{s_b}$, yields the E phase with residual symmetry
$U(1)\times U(1)$; (ii) $\phi^-\neq \phi^+=0$, implying $\varphi_{c_a}-\varphi_{c_b}
-\varphi_{s_a}-\varphi_{s_b}=0$, yields the C phase with residual symmetry
$U(1)\times U(1)\times U(1)$; (iii) $|\phi^+|=|\phi^-|\neq 0$, 
implying $\varphi_{c_a}=\varphi_{c_b}$ and $\varphi_{s_a}=-\varphi_{s_b}$, 
yields the L phase with residual symmetry $U(1)\times [U(1) \rtimes \mathbb{Z}_2]$. 
The $\mathbb{Z}_2$ symmetry of the L phase arises from a $\pi$-spin-rotation
along the axis parallel to $\mathbf{\phi}$. While 'direction of $\mathbf{\phi}$'
is not well defined in general as one needs two real vectors to capture the information
contained in $\mathbf{\phi}$, the property $|\phi^+|=|\phi^-|$ of the L phase
implies that $\mathbf{\phi}$ can be written as a real vector times a phase factor.

The effects of an external magnetic field along the $z$-axis and the cross-hopping are easily
seen. With non-zero external field the $\mathbb{Z}_2$ factor in $\Gamma$ disappears and thus
the L and E phase are not distinguished anymore. With non-zero cross-hopping 
the symmetry of the Hamiltonian $\Gamma$ reduces to $U(1)\times [U(1) \rtimes \mathbb{Z}_2]$ 
(corresponding to applying a condition $\varphi_{c_a}=\varphi_{c_b}$ 
and $\varphi_{s_a}=\varphi_{s_b}$). As a result the C and E phases are not 
distinguished any more.

\subsection{Rotationally invariant interaction}
\label{sec:su2}
While the presented calculations were performed with density-density interaction 
$H^{\text{dd}}_{\text{int}}$ only, real materials usually possess the full
spin rotational invariance, which is restored by the first term of $H'_{\text{int}}$.
It is therefore important to show that including $H'_{\text{int}}$ does not lead
to qualitatively different results. Considering $H'_{\text{int}}$ we have to distinguish
the case with ($J'\neq0$) and without ($J'=0$) pair-hopping, which breaks the 
separate conservations of $a$ and $b$ charge. The symmetry of Hamiltonian
(\ref{eq:hubbard}) with $H'_{\text{int}}$ is $U(1)\times SU(2)$ in the former case
and $U(1)\times U(1)\times SU(2)$ in the latter one. We will consider the latter case
and show that it leads to the same phases as the density-density interaction with no
cross-hopping.

First, we discuss the Landau functional.
Treating $\boldsymbol{\phi}$ as
a 3-dimensional complex vector the Landau functional must be a rotationally 
invariant function of two vectors $F(\bar{\boldsymbol{\phi}},\boldsymbol{\phi})$, which
implies that its Taylor expansion must be a polynomial in 
$\bar{\boldsymbol{\phi}}\cdot\bar{\boldsymbol{\phi}}$,
$\boldsymbol{\phi}\cdot\boldsymbol{\phi}$ and $\bar{\boldsymbol{\phi}}\cdot\boldsymbol{\phi}$.~\cite{note3}
Moreover, $F(\bar{\boldsymbol{\phi}},\boldsymbol{\phi})$ must be invariant 
with respect to the overall phase of $\boldsymbol{\phi}$ (due to
$V_{1,2}=0$ and $J'=0$). This reduces the expansion of 
$F(\bar{\boldsymbol{\phi}},\boldsymbol{\phi})$ to
a polynomial in $\bar{\boldsymbol{\phi}}\cdot\boldsymbol{\phi}$ and 
$|\bar{\boldsymbol{\phi}}\wedge\boldsymbol{\phi}|^2$. With the $z$-axis parallel
to $i(\bar{\boldsymbol{\phi}}\wedge\boldsymbol{\phi})$ (or any normal
to $\boldsymbol{\phi}$ in case the wedge product is zero) 
the two terms yield the $|\phi^+|^2+|\phi^-|^2$ and 
$(|\phi^+|^2-|\phi^-|^2)^2$ and the equivalence to 
(\ref{eq:landau1}) becomes explicit. 

Using the same choice of the coordinates we can repeat the discussion from the previous 
section. For $|\phi^+|\neq|\phi^-|$ only rotations about the $z$-axis can preserve 
$\boldsymbol{\phi}$ and thus we can use expression (\ref{eq:transf}). However, 
since the spin-flip term does not allow independent spin rotations about the $z$-axis
we have to assume $\varphi_{s_a}=\varphi_{s_b}\equiv\varphi_s$ from the beginning.
The symmetries of the E phase thus must fulfill $\varphi_s=0$ and 
$\varphi_{c_a}=\varphi_{c_b}$, i.e., only $U(1)$ symmetry corresponding 
to the conservation of total charge is preserved. The invariance of $\boldsymbol{\phi}$
in the C phase is less restrictive. The condition $\varphi_{c_a}-\varphi_{c_b}-\varphi_s=0$
leads to residual symmetry $U(1)\times U(1)$ allowing also spin rotations
accompanied by change of the relative phase of the $a$ and $b$ orbitals.
As in the density-density case, non-zero cross-hopping or pair-hopping removes
the distinction between the C and E phases. Finally, if 
$\bar{\boldsymbol{\phi}}\wedge\boldsymbol{\phi}=0$ we can choose the $z$-axis parallel
to $\bar{\boldsymbol{\phi}}+\boldsymbol{\phi}$, in which case the only non-zero
component of $\boldsymbol{\phi}$, $\phi_z\sim a^{\dagger}_{\uparrow}b^{\phantom\dagger}_{
\uparrow}-a^{\dagger}_{\downarrow}b^{\phantom\dagger}_{\downarrow}$, is invariant
under $U(1)\times U(1)$ corresponding to $z-axis$ spin rotation (besides the overall
phase change), which correspond to the $\mathbb{Z}_2$ symmetry of the L phase 
in the density-density case. Note, that the non-trivial $U(1)$ symmetries 
of the C and L phase are different. While the former one contains spin-rotations
about an axis perpendicular to $\boldsymbol{\phi}$ combined with orbital phase 
transformations, the latter one consists of pure spin-rotations about 
an axis 'parallel' to $\boldsymbol{\phi}$.

While the same excitonic phases exist in the model with and without $H'_{\text{int}}$ 
the space of low-energy fluctuations of the order parameter, which reflects
the broken symmetries, will be quite different in the two cases. Since our numerical
method does not treat these fluctuations we do not discuss them here.

\section{Discussion}
The above numerical results have demonstrated the rich phase diagram of the excitonic
condensate. The key feature allowing the various phases is the orbital-off-diagonal 
character, which renders the order parameter complex. Were the order parameter real,
such as the usual spontaneous magnetization, only one phase would have been possible since
all real magnetization vectors of the same magnitude can be connected by some spin rotation (This does exclude
multiple phases with different translational symmetry, e.g., ferro and antiferro).
This is not true for complex vectors. To see this, one can consider the vector
product $\bar{\boldsymbol{\phi}}\wedge\boldsymbol{\phi}$, which is invariant under rotations
as well as overall phase transformation. Two states characterized by order parameters
with the same magnitude $|\boldsymbol{\phi}|^2$ but different $\bar{\boldsymbol{\phi}}\wedge\boldsymbol{\phi}$
cannot be transformed into one another by a symmetry operation of the Hamiltonian.
It is easy to see that the L phase corresponds to $\bar{\boldsymbol{\phi}}\wedge\boldsymbol{\phi}=0$,
while for the C phase $|\bar{\boldsymbol{\phi}}\wedge\boldsymbol{\phi}|=|\boldsymbol{\phi}|^2$.

How do the different EC states look like in the direct space?
Halperin and Rice~\cite{halperin68b} have shown that exciton condensation
in the spin-triplet channel gives rise either to a spin-density or
a spin-current-density wave. Assuming real Wannier orbitals $\psi_a(\mathbf{r})$ and $\psi_b(\mathbf{r})$, 
and neglecting (although this is not essential) the overlap density of orbitals on different sites,
the exciton condensation gives rise to the spin-density $S^{\alpha}(\mathbf{r})$ and the spin-current density
$J^{\alpha\beta}(\mathbf{r})$~\cite{balents00b}
\begin{equation}
	\label{eq:dnst}
	\begin{split}
		S^{\alpha}(\mathbf{r})&=
		2\psi_a(\mathbf{r})\psi_b(\mathbf{r})\operatorname{Re}\phi^{\alpha}\\  
		J^{\alpha\beta}(\mathbf{r})&=
		\left(\psi_a(\mathbf{r})\partial_{\beta}\psi_b(\mathbf{r})
		-\psi_a(\mathbf{r})\partial_{\beta}\psi_b(\mathbf{r})\right)\operatorname{Im}\phi^{\alpha}.
	\end{split}
\end{equation}
on each lattice site. In our model with $H^{\text{dd}}_{\text{int}}$ only, 
the spin polarization is confined to $x$--$y$ plane.
Note that the total integrated spin moment or integrated spin current in (\ref{eq:dnst}) are zero.
Real $\phi$ thus gives rise to magnetic multipoles, precise form of which depends
on the shapes of orbital $\psi_a(\mathbf{r})$ and $\psi_b(\mathbf{r})$. However, since the low lying
excitations in general do not correspond to direct space rotations of these multipoles, 
we do not find it useful to speak about multipole order. 
In fact, in the present case with $V_{1,2}=0$ a
spin-density state can be continuously 'rotated' into a spin-current state, by varying the phase of $\boldsymbol{\phi}$,
without any energy cost.
In the L phase, all Cartesian components of $\boldsymbol{\phi}$ can be made simultaneously real or imaginary,
thus corresponding to a purely spin-density or a spin-current state. Non-zero cross hopping $V_{1,2}$
removes the degeneracy of spin-density and spin-current states.~\cite{balents00a,kunes14a}

In the C and E phases, on the other hand, a finite phase difference between 
the Cartesian components of $\boldsymbol{\phi}$ implies that
both the spin density and spin current must be simultaneously present. 
In these phases, breaking of the $\mathbb{Z}_2$ symmetry in the Hamiltonian
with $H^{\text{dd}}_{\text{int}}$ only gives rise to a 
net spin moment along the $z$-axis. In case with $SU(2)$ symmetry of the Coulomb interaction,
$\boldsymbol{\phi}$ can point in arbitrary direction and the net spin moment
is parallel to $i\bar{\boldsymbol{\phi}}\wedge\boldsymbol{\phi}$. As discussed in detail
by Balents~\cite{balents00a} this ferromagnetic phase is distinct from the excitonic
ferromagnetism proposed by Volkov~{\it et al.}~\cite{volkov75} which arises from mixing
of spin-singlet and spin-triplet orders. Such mixing is possible only for small exchange $J$. 

The present ferromagnetic order is induced by the exciton condensation 
in the sense that the normal phase is not close to a ferromagnetic instability.
The magnitude of the ordered moment is very small compared to the transition temperature,
given there is no geometric frustration in the system.
The magnetization close to the N/C boundary follows a linear $1-\tfrac{T}{T_c}$ dependence
rather than the square root $(1-\tfrac{T}{T_c})^{1/2}$ behavior expected of the mean-field order
parameter, obeyed by $\phi^-$. This behavior is easily understood in both weak and strong coupling
mean-field theories which give $\langle m_z\rangle\sim|\phi^-|^2$ in the leading
order of the perturbation theory in $\phi^-$.
\begin{figure}
\includegraphics[width=0.5\columnwidth,angle=270,clip]{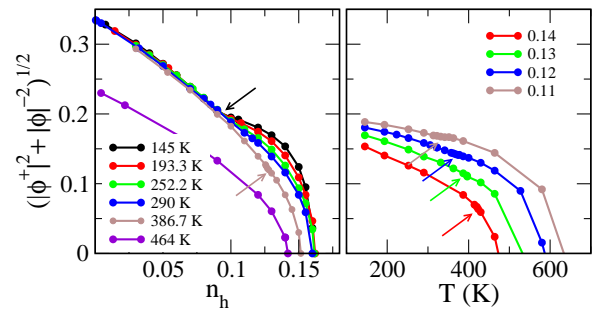}
\caption{ \label{fig:phi2}
	The magnitude of the order parameter $\sqrt{|\phi^+|^2+|\phi^-|^2}$. Along the same 
	constant $T$ (right) and constant $n_h$ (left) scans as in Figs.~\ref{fig:T-cut} and
	\ref{fig:n-cut}. The arrows mark the narrow region of the E phase.
}
\end{figure}
\begin{figure}
\includegraphics[width=0.7\columnwidth,angle=270,clip]{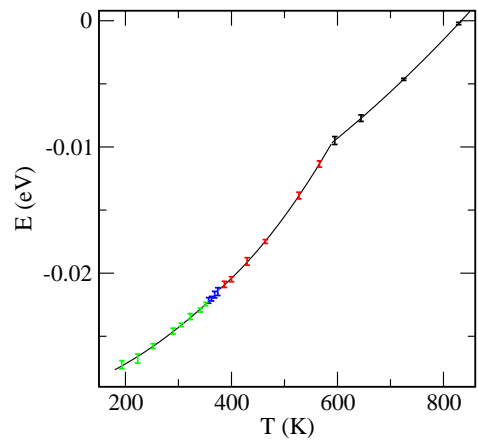}
\caption{ \label{fig:etot}
The internal energy per site as a function of temperature along the $n_h=0.12$
scan through the phase diagram. The colors of the error bars code the 
phases as in Fig.~\ref{fig:phase_diag}. The lines represent quadratic fits to 
the N, L and C phases.
	}
\end{figure}

The fundamental nature of $\boldsymbol{\phi}$ as an order parameter can be also seen in Fig.~\ref{fig:phi2}
where we plot the magnitude $|\boldsymbol{\phi}|^2$. Both the constant--$n_h$ and constant--$T$
scans show that the L--E--C transition can be viewed as a rotation of the 
vector $(|\phi^+|,|\phi^-|)$. This is particularly so in the vicinity of the multi-critical point.
This observation shows that our numerical data are consistent with the Landau theory and
that at the multi-critical point the system has higher symmetry than the Hamiltonian (\ref{eq:hubbard}).

What determines the stability of the different phases on the microscopic level?
Close to the multi-critical point one can expect separation of energy scales 
responsible for the condensation itself and for selecting of a particular phase.
This separation is clearly visible in Fig.~\ref{fig:etot}
where the internal energy across all the phases is shown along a constant--$n_h=0.12$
scan. While there is a distinct kink at 790~K where the exciton condensation takes place,
the L/E and E/C transitions are hardly discernible within our numerical accuracy.

It is not possible to link the different excitonic phases to 
particular terms in Hamiltonian (\ref{eq:hubbard}) in general. It is, nevertheless, instructive 
to discuss the strong and the weak coupling limits even though, given 
we are dealing with doped systems, they may provide only a qualitative insight.
The strong-coupling Hamiltonian of Ref.~\onlinecite{kunes14a}
\begin{equation}
\label{eq:boson1}
\begin{split}
H^{\text{dd}}_{\text{eff}}&=\sum_{i} \mu n_i+K_{\perp}\sum_{ij,s}d^{\dagger}_{i,s}d^{\phantom\dagger}_{j,s}
  +\sum_{\langle ij \rangle}\bigl(K_{\parallel}n_in_j
   + K_0S^z_i S^z_j\bigr) \\
  &+K_1\sum_{\langle ij \rangle,s} \left(
   d^{\dagger}_{i,s}d^{\dagger}_{j,-s}+d^{\phantom\dagger}_{i,s}
  d^{\phantom\dagger}_{j,-s}\right)
\end{split}
\end{equation}
describes two flavors $s=\pm1$ of bosons with the hard-core constraint
$n_i=\sum_{s}d^{\dagger}_{i,s}d^{\phantom\dagger}_{i,s}\le 1$,
corresponding to HS states created by
$d^{\dagger}_{1}=a^{\dagger}_{\uparrow} b^{\phantom\dagger}_{\downarrow}$ and
$d^{\dagger}_{-1}=a^{\dagger}_{\downarrow}b^{\phantom\dagger}_{\uparrow}$ out
of the LS vacuum. Responsible for condensation of the excitons is the hopping term
proportional to $K_{\perp}$. This term, however, has a higher symmetry than Hamiltonian
(\ref{eq:hubbard}) and does not distinguish between excitonic phases. 
The stable phase is therefore selected by the $K_0$ and $K_1$ terms. If negative the spin exchange term
$K_0$ favors the ferromagnetic C phase, while positive~\cite{note2}
$K_0$ favors the unpolarized L
phase. The $K_1$ term, present only for $V_{1,2}\neq0$, yields zero 
contribution to $\langle d^{\phantom\dagger}_{i,s}\rangle\langle
d^{\phantom\dagger}_{j,-s}\rangle$ in the C phase, while in the L phase 
this contribution can be made negative by proper choice of the 
phase of $\langle d^{\phantom\dagger}_{i,s}\rangle$. The $K_1$ term
thus favors the L phase. Strong coupling analysis of the case $J\ll K_0,K_{\perp}$,
which requires inclustion of the spin-singlet bosons, can be
found in Ref.~\onlinecite{balents00a}.

In the weak coupling limit the excitonic condensation is driven by the 
$(U-2J)n^a_{i\sigma}n^b_{i-\sigma}$ term in (\ref{eq:hubbard}).
The repulsion between electrons in the same orbital $Un^a_{i\uparrow}n^a_{i\downarrow}$
favors the polarized C phase over L phase since the expectation value
$\langle n^a_{i\uparrow}n^a_{i\downarrow} \rangle$ vanishes in the former.

We close the discussion by a technical remark. While the exciton condensation is obtained
also in a static Hartree theory, the full treatment of the local dynamics is crucial
to obtain the different excitonic phases. Our brief experimenting with Hartree theory
revealed a strong bias towards the polarized C phase, which can be traced back to 
the minimization of $\langle n^a_{i\uparrow}n^a_{i\downarrow} \rangle$.
In a static theory this is achieved by polarizing the system while in the dynamic theory
these terms are strongly suppressed already in the normal phase.

\section{Conclusions}
Using dynamical mean-field theory
we have investigated phase diagram of two-band Hubbard model with strong Hund's
coupling in the regime of excitonic instability. The spin-triplet orbital-off-diagonal
order parameter allows several thermodynamic phases. 
Varying temperature and the chemical potential of the system we have observed all 
the symmetry allowed phases including those with induced ferromagnetic polarization.
The transitions at low temperatures are of the first order accompanied 
by a charge separation. At intermediate temperatures, however, the transitions 
become continuous. The ferromagnetic polarization exhibits a large reversible
response to small changes of the chemical potential in this regime.
This effect can be used to construct devices where magnetization is 
switched reversibly by means of a gate voltage.

\begin{acknowledgements}
We thank P. Augustinsk\'y, A. Kauch, V. Pokorn\'y, D. Vollhardt, and A.~P. Kampf for discussions
and valuable suggestions. We acknowledge the support of Deutsche Forschungsgemeinschaft through FOR1346
and the Grant Agency of the Czech Republic through project 13-25251S.
\end{acknowledgements}

\appendix
\section{}
\label{sec:a1}
In order to analyze the behavior of (\ref{eq:landau1}) in the vicinity of the multi-critical
point we write the functional to the 8th order in $\phi$ and use polar coordinates
in the $|\phi^+|$--$|\phi^-|$ plane:
\begin{equation}
\label{eq:landau}
\begin{split}
   F(\eta,t)=\alpha\eta+\beta_0\eta^2+\gamma_0\eta^3+\delta_0\eta^4+ \\
      \left(\beta_1\eta^2+\gamma_1\eta^3+\delta_1\eta^4\right)C(t)^2+\delta_2\eta^4C(t)^4 \\
     |\phi^+|=\eta^{\frac{1}{2}}\cos(t),\quad |\phi^-|=\eta^{\frac{1}{2}}\sin(t),
    \end{split}
      \end{equation}
where $C(t)=\cos(2t)$. For sufficiently small $\alpha$ and $\beta_1$, $F(\eta,t)$ has
approximately the form of a Mexican hat with the radial minima $\tilde{\eta}(t)$ on a
circle $\tilde{\eta}(t)=-\tfrac{\alpha}{2\beta_0}$. The terms containing $C(t)$ 
cause a deformation of $\tilde{\eta}(t)$ from the circular shape and lift the degeneracy
of $F$ along $\tilde{\eta}(t)$. Our strategy will be to analyze this effect in different orders of
$\alpha$. Behavior of the stationary points of $F(\eta,t)$ depends on the direction in which the limit
$\beta_1\rightarrow0, \alpha\rightarrow0^-$ is taken. We specify the limit by introducing
a parameter $\Lambda$: $\beta_1= \Lambda\alpha$. 

The variation of Landau functional along the contour of radial minima $\tilde{\eta}(t)$
is given by $\tilde{F}(t)\equiv F(\tilde{\eta}(t),t)$. The $\tilde{\eta}(t)$
itself fulfills the equation 
\begin{equation}
	\label {eq:radial}
        \partial_{\eta}F(\tilde{\eta}(t),t)=0.
\end{equation}
Solving (\ref{eq:radial}) in the form of a power series $\tilde{\eta}(t)=K(t)\alpha+L(t)\alpha^2+M(t)\alpha^3$ 
we get
\begin{equation}
 \begin{split}
K(t)&=-\frac{1}{2\beta_0}\\
L(t)&=\frac{1}{8\beta_0^3} \left(-3\gamma_0+4\Lambda\beta_0C(t)^2-3\gamma_1C(t)^2\right)\\
M(t)&=\frac{1}{16\beta_0^5}\left(-9\gamma_0+4\beta_0\delta_0+18\Lambda\beta_0\gamma_0C(t)^2
18\gamma_0\gamma_1C(t)^2+\right.\\
        &\left.+4\beta_0\delta_1C(t)^2-4\Lambda^2\beta_0^2C(t)^4-9\gamma_1^2C(t)^4+4\beta_0\delta_2C(t)^4\right)
        \end{split}.
\end{equation}
Substituting this expansion into (\ref{eq:landau}) we arrive at the expansion of $\tilde{F}(t)$
\begin{equation}
 \label{eq:expand}
 \begin{split}
   \tilde{F}(t)&=\frac{2\Lambda\beta_0-\gamma_1}{8\beta_0^3}\alpha^3C(t)^2\\
               &+\frac{2\beta_0\left(6\Lambda\gamma_0+\delta_1\right)-9\gamma_0\gamma_1}{32\beta_0^5}\alpha^4C(t)^2\\
&-\frac{16\Lambda^2\beta_0^2+9\gamma_1^2-
   4\beta_0\left(6\Lambda\gamma_1+\delta_2\right)}{64\beta_0^5}\alpha^4C(t)^4+o(\alpha^5),
\end{split}
\end{equation}
where we have dropped the $t$-constant terms. For a general $\Lambda$, $C(t)^2$ 
appears in the order $\alpha^3$ while $C(t)^4$ appears first in the order $\alpha^4$.
Therefore, with the exception of $\Lambda=\tfrac{\gamma_0}{2\beta_0}$ where the $\alpha^3$ term
in (\ref{eq:expand}) vanishes, the $C(t)^2$ term is dominant for a sufficiently small $\alpha$. 
$C(t)^2$ has a unique minimum either at $t=0$ or at $t=\pi/4$, depending on the sign
of the prefactor. The functional (\ref{eq:landau}) in this limit thus yields a single stable phase 
- the C phase ($t=0$) for $2\Lambda\beta_0-\gamma_1<0$ and L phase for $2\Lambda\beta_0-\gamma_1>0$.
Varying $\Lambda$ across the $\Lambda=\tfrac{\gamma_0}{2\beta_0}$ the minimum and maximum
of $\tilde{F}(t)$ switch and there is a transition between the C and L phases.

This transition can happen in one of two distinct ways, see Fig.~\ref{fig:cartoon}. 
To see this we keep $\alpha$ fixed
(and sufficiently small so that all $o(\alpha^5)$ terms in $\tilde{F}(t)$ are irrelevant).
In a narrow interval of $\Lambda$ (of the width of order $\alpha$) in the vicinity of
$\Lambda=\tfrac{\gamma_0}{2\beta_0}$ the $\alpha^3$ contribution to $\tilde{F}(t)$ vanishes and
$\tilde{F}(t)$ has the form $aC(t)^2+bC(t)^4$ with coefficients $a$ and $b$ of comparable sizes.
This angular form offers only three possibilities for its stationary points:
i) the only stationary points are at $t=0$ and $\pi/4$,
in which case one of them must be maximum and the other one minimum corresponding to a single stable phase,
either L or C, ii) there is a single local maximum between 0 and $\pi/4$, in which case there
must be local minima at 0 and $\pi/4$ corresponding to L and C phases being simultaneously stable,
iii) there is a single local minimum between 0 and $\pi/4$ corresponding to the E phase, in which case there
must be local maxima at 0 and $\pi/4$ and thus E being the only stable phase.
While (i) includes the case where the $C(t)^2$ term is dominant, the transition between the C and L
can only take place by passing through (ii), i.e., a first order transition, or (iii), i.e., two
continuous transitions with an intermediate E phase.


\begin{thebibliography}{34}
\expandafter\ifx\csname natexlab\endcsname\relax\def\natexlab#1{#1}\fi
\expandafter\ifx\csname bibnamefont\endcsname\relax
  \def\bibnamefont#1{#1}\fi
\expandafter\ifx\csname bibfnamefont\endcsname\relax
  \def\bibfnamefont#1{#1}\fi
\expandafter\ifx\csname citenamefont\endcsname\relax
  \def\citenamefont#1{#1}\fi
\expandafter\ifx\csname url\endcsname\relax
  \def\url#1{\texttt{#1}}\fi
\expandafter\ifx\csname urlprefix\endcsname\relax\def\urlprefix{URL }\fi
\providecommand{\bibinfo}[2]{#2}
\providecommand{\eprint}[2][]{\url{#2}}

\bibitem[{\citenamefont{Vollhardt and W\"olfle}(2013)}]{vollhardt}
\bibinfo{author}{\bibfnamefont{D.}~\bibnamefont{Vollhardt}} \bibnamefont{and}
  \bibinfo{author}{\bibfnamefont{P.}~\bibnamefont{W\"olfle}},
  \emph{\bibinfo{title}{The Superfluid Phases of Helium 3}}
  (\bibinfo{publisher}{Dover Publications}, \bibinfo{address}{Reprint edition},
  \bibinfo{year}{2013}).

\bibitem[{\citenamefont{Mott}(1961)}]{mott61}
\bibinfo{author}{\bibfnamefont{N.}~\bibnamefont{Mott}},
  \bibinfo{journal}{Philos. Mag.} \textbf{\bibinfo{volume}{6}},
  \bibinfo{pages}{287} (\bibinfo{year}{1961}).

\bibitem[{\citenamefont{Halperin and Rice}(1968{\natexlab{a}})}]{halperin68a}
\bibinfo{author}{\bibfnamefont{B.~I.} \bibnamefont{Halperin}} \bibnamefont{and}
  \bibinfo{author}{\bibfnamefont{T.~M.} \bibnamefont{Rice}},
  \bibinfo{journal}{Rev. Mod. Phys.} \textbf{\bibinfo{volume}{40}},
  \bibinfo{pages}{755} (\bibinfo{year}{1968}{\natexlab{a}}).

\bibitem[{\citenamefont{Halperin and Rice}(1968{\natexlab{b}})}]{halperin68b}
\bibinfo{author}{\bibfnamefont{B.}~\bibnamefont{Halperin}} \bibnamefont{and}
  \bibinfo{author}{\bibfnamefont{T.~M.} \bibnamefont{Rice}},
  \bibinfo{journal}{Solid State Physics} \textbf{\bibinfo{volume}{21}},
  \bibinfo{pages}{115} (\bibinfo{year}{1968}{\natexlab{b}}).

\bibitem[{\citenamefont{Balents and Varma}(2000)}]{balents00a}
\bibinfo{author}{\bibfnamefont{L.}~\bibnamefont{Balents}} \bibnamefont{and}
  \bibinfo{author}{\bibfnamefont{C.~M.} \bibnamefont{Varma}},
  \bibinfo{journal}{Phys. Rev. Lett.} \textbf{\bibinfo{volume}{84}},
  \bibinfo{pages}{1264} (\bibinfo{year}{2000}).

\bibitem[{\citenamefont{Balents}(2000)}]{balents00b}
\bibinfo{author}{\bibfnamefont{L.}~\bibnamefont{Balents}},
  \bibinfo{journal}{Phys. Rev. B} \textbf{\bibinfo{volume}{62}},
  \bibinfo{pages}{2346} (\bibinfo{year}{2000}).

\bibitem[{\citenamefont{Barzykin and Gor'kov}(2000)}]{barzykin00}
\bibinfo{author}{\bibfnamefont{V.}~\bibnamefont{Barzykin}} \bibnamefont{and}
  \bibinfo{author}{\bibfnamefont{L.~P.} \bibnamefont{Gor'kov}},
  \bibinfo{journal}{Phys. Rev. Lett.} \textbf{\bibinfo{volume}{84}},
  \bibinfo{pages}{2207} (\bibinfo{year}{2000}).

\bibitem[{\citenamefont{Veillette and Balents}(2001)}]{veillette01}
\bibinfo{author}{\bibfnamefont{M.~Y.} \bibnamefont{Veillette}}
  \bibnamefont{and} \bibinfo{author}{\bibfnamefont{L.}~\bibnamefont{Balents}},
  \bibinfo{journal}{Phys. Rev. B} \textbf{\bibinfo{volume}{65}},
  \bibinfo{pages}{014428} (\bibinfo{year}{2001}).

\bibitem[{\citenamefont{Eisenstein and MacDonald}(2004)}]{eisenstein04}
\bibinfo{author}{\bibfnamefont{J.}~\bibnamefont{Eisenstein}} \bibnamefont{and}
  \bibinfo{author}{\bibfnamefont{A.~H.} \bibnamefont{MacDonald}},
  \bibinfo{journal}{Nature} \textbf{\bibinfo{volume}{432}},
  \bibinfo{pages}{691} (\bibinfo{year}{2004}).

\bibitem[{\citenamefont{Rademaker
  et~al.}(2013{\natexlab{a}})\citenamefont{Rademaker, van~den Brink, Zaanen,
  and Hilgenkamp}}]{rademaker13a}
\bibinfo{author}{\bibfnamefont{L.}~\bibnamefont{Rademaker}},
  \bibinfo{author}{\bibfnamefont{J.}~\bibnamefont{van~den Brink}},
  \bibinfo{author}{\bibfnamefont{J.}~\bibnamefont{Zaanen}}, \bibnamefont{and}
  \bibinfo{author}{\bibfnamefont{H.}~\bibnamefont{Hilgenkamp}},
  \bibinfo{journal}{Phys. Rev. B} \textbf{\bibinfo{volume}{88}},
  \bibinfo{pages}{235127} (\bibinfo{year}{2013}{\natexlab{a}}).

\bibitem[{\citenamefont{Rademaker
  et~al.}(2013{\natexlab{b}})\citenamefont{Rademaker, Johnston, Zaanen, and
  van~den Brink}}]{rademaker13b}
\bibinfo{author}{\bibfnamefont{L.}~\bibnamefont{Rademaker}},
  \bibinfo{author}{\bibfnamefont{S.}~\bibnamefont{Johnston}},
  \bibinfo{author}{\bibfnamefont{J.}~\bibnamefont{Zaanen}}, \bibnamefont{and}
  \bibinfo{author}{\bibfnamefont{J.}~\bibnamefont{van~den Brink}},
  \bibinfo{journal}{Phys. Rev. B} \textbf{\bibinfo{volume}{88}},
  \bibinfo{pages}{235115} (\bibinfo{year}{2013}{\natexlab{b}}).

\bibitem[{\citenamefont{Khaliullin}(2013)}]{khaliullin13}
\bibinfo{author}{\bibfnamefont{G.}~\bibnamefont{Khaliullin}},
  \bibinfo{journal}{Phys. Rev. Lett.} \textbf{\bibinfo{volume}{111}},
  \bibinfo{pages}{197201} (\bibinfo{year}{2013}).

\bibitem[{\citenamefont{Kune\v{s} and Augustinsk\'y}()}]{kunes14b}
\bibinfo{author}{\bibfnamefont{J.}~\bibnamefont{Kune\v{s}}} \bibnamefont{and}
  \bibinfo{author}{\bibfnamefont{P.}~\bibnamefont{Augustinsk\'y}},
  \bibinfo{note}{arXiv:1405.1191}.

\bibitem[{\citenamefont{Volkov et~al.}(1975)\citenamefont{Volkov, Kopaev, and
  Rusinov}}]{volkov75}
\bibinfo{author}{\bibfnamefont{B.~A.} \bibnamefont{Volkov}},
  \bibinfo{author}{\bibfnamefont{Y.~V.} \bibnamefont{Kopaev}},
  \bibnamefont{and} \bibinfo{author}{\bibfnamefont{A.~I.}
  \bibnamefont{Rusinov}}, \bibinfo{journal}{JETP}
  \textbf{\bibinfo{volume}{41}}, \bibinfo{pages}{952} (\bibinfo{year}{1975}).

\bibitem[{\citenamefont{des Cloizeaux}(1965)}]{descloizeaux65}
\bibinfo{author}{\bibfnamefont{J.}~\bibnamefont{des Cloizeaux}},
  \bibinfo{journal}{J. Phys. Chem. Solids} \textbf{\bibinfo{volume}{26}},
  \bibinfo{pages}{259} (\bibinfo{year}{1965}).

\bibitem[{\citenamefont{Zocher et~al.}(2011)\citenamefont{Zocher, Timm, and
  Brydon}}]{zocher11}
\bibinfo{author}{\bibfnamefont{B.}~\bibnamefont{Zocher}},
  \bibinfo{author}{\bibfnamefont{C.}~\bibnamefont{Timm}}, \bibnamefont{and}
  \bibinfo{author}{\bibfnamefont{P.~M.~R.} \bibnamefont{Brydon}},
  \bibinfo{journal}{Phys. Rev. B} \textbf{\bibinfo{volume}{84}},
  \bibinfo{pages}{144425} (\bibinfo{year}{2011}).

\bibitem[{\citenamefont{Sachdev and Bhatt}(1990)}]{sachdev90}
\bibinfo{author}{\bibfnamefont{S.}~\bibnamefont{Sachdev}} \bibnamefont{and}
  \bibinfo{author}{\bibfnamefont{R.~N.} \bibnamefont{Bhatt}},
  \bibinfo{journal}{Phys. Rev. B} \textbf{\bibinfo{volume}{41}},
  \bibinfo{pages}{9323} (\bibinfo{year}{1990}).

\bibitem[{\citenamefont{Sommer et~al.}(2001)\citenamefont{Sommer, Vojta, and
  Becker}}]{sommer01}
\bibinfo{author}{\bibfnamefont{T.}~\bibnamefont{Sommer}},
  \bibinfo{author}{\bibfnamefont{M.}~\bibnamefont{Vojta}}, \bibnamefont{and}
  \bibinfo{author}{\bibfnamefont{K.~W.} \bibnamefont{Becker}},
  \bibinfo{journal}{Eur. Phys. J. B} \textbf{\bibinfo{volume}{23}},
  \bibinfo{pages}{329} (\bibinfo{year}{2001}).

\bibitem[{\citenamefont{Giamarchi et~al.}(2008)\citenamefont{Giamarchi,
  R\"uegg, and Tchernyschyov}}]{giamarchi08}
\bibinfo{author}{\bibfnamefont{T.}~\bibnamefont{Giamarchi}},
  \bibinfo{author}{\bibfnamefont{C.}~\bibnamefont{R\"uegg}}, \bibnamefont{and}
  \bibinfo{author}{\bibfnamefont{O.}~\bibnamefont{Tchernyschyov}},
  \bibinfo{journal}{Nat. Phys.} \textbf{\bibinfo{volume}{4}},
  \bibinfo{pages}{198} (\bibinfo{year}{2008}).

\bibitem[{\citenamefont{Kaneko et~al.}(2012)\citenamefont{Kaneko, Seki, and
  Ohta}}]{kaneko12}
\bibinfo{author}{\bibfnamefont{T.}~\bibnamefont{Kaneko}},
  \bibinfo{author}{\bibfnamefont{K.}~\bibnamefont{Seki}}, \bibnamefont{and}
  \bibinfo{author}{\bibfnamefont{Y.}~\bibnamefont{Ohta}},
  \bibinfo{journal}{Phys. Rev. B} \textbf{\bibinfo{volume}{85}},
  \bibinfo{pages}{165135} (\bibinfo{year}{2012}).

\bibitem[{\citenamefont{Kune\ifmmode~\check{s}\else \v{s}\fi{} and
  Augustinsk\'y}(2014)}]{kunes14a}
\bibinfo{author}{\bibfnamefont{J.}~\bibnamefont{Kune\ifmmode~\check{s}\else
  \v{s}\fi{}}} \bibnamefont{and}
  \bibinfo{author}{\bibfnamefont{P.}~\bibnamefont{Augustinsk\'y}},
  \bibinfo{journal}{Phys. Rev. B} \textbf{\bibinfo{volume}{89}},
  \bibinfo{pages}{115134} (\bibinfo{year}{2014}).

\bibitem[{\citenamefont{Kaneko and Ohta}()}]{kaneko14}
\bibinfo{author}{\bibfnamefont{T.}~\bibnamefont{Kaneko}} \bibnamefont{and}
  \bibinfo{author}{\bibfnamefont{Y.}~\bibnamefont{Ohta}},
  \bibinfo{note}{arXiv:1407.4872}.

\bibitem[{\citenamefont{Werner and Millis}(2007)}]{werner07}
\bibinfo{author}{\bibfnamefont{P.}~\bibnamefont{Werner}} \bibnamefont{and}
  \bibinfo{author}{\bibfnamefont{A.~J.} \bibnamefont{Millis}},
  \bibinfo{journal}{Phys. Rev. Lett.} \textbf{\bibinfo{volume}{99}},
  \bibinfo{pages}{126405} (\bibinfo{year}{2007}).

\bibitem[{\citenamefont{Suzuki et~al.}(2009)\citenamefont{Suzuki, Watanabe, and
  Ishihara}}]{suzuki09}
\bibinfo{author}{\bibfnamefont{R.}~\bibnamefont{Suzuki}},
  \bibinfo{author}{\bibfnamefont{T.}~\bibnamefont{Watanabe}}, \bibnamefont{and}
  \bibinfo{author}{\bibfnamefont{S.}~\bibnamefont{Ishihara}},
  \bibinfo{journal}{Phys. Rev. B} \textbf{\bibinfo{volume}{80}},
  \bibinfo{pages}{054410} (\bibinfo{year}{2009}).

\bibitem[{\citenamefont{Blume et~al.}(1971)\citenamefont{Blume, Emery, and
  Griffiths}}]{beg}
\bibinfo{author}{\bibfnamefont{M.}~\bibnamefont{Blume}},
  \bibinfo{author}{\bibfnamefont{V.~J.} \bibnamefont{Emery}}, \bibnamefont{and}
  \bibinfo{author}{\bibfnamefont{R.~B.} \bibnamefont{Griffiths}},
  \bibinfo{journal}{Phys. Rev. A} \textbf{\bibinfo{volume}{4}},
  \bibinfo{pages}{1071} (\bibinfo{year}{1971}).

\bibitem[{\citenamefont{Kune\ifmmode~\check{s}\else \v{s}\fi{} and
  K\ifmmode~\check{r}\else \v{r}\fi{}\'apek}(2011)}]{kunes11}
\bibinfo{author}{\bibfnamefont{J.}~\bibnamefont{Kune\ifmmode~\check{s}\else
  \v{s}\fi{}}} \bibnamefont{and}
  \bibinfo{author}{\bibfnamefont{V.}~\bibnamefont{K\ifmmode~\check{r}\else
  \v{r}\fi{}\'apek}}, \bibinfo{journal}{Phys. Rev. Lett.}
  \textbf{\bibinfo{volume}{106}}, \bibinfo{pages}{256401}
  (\bibinfo{year}{2011}).

\bibitem[{\citenamefont{Georges et~al.}(1996)\citenamefont{Georges, Kotliar,
  Krauth, and Rozenberg}}]{dmft}
\bibinfo{author}{\bibfnamefont{A.}~\bibnamefont{Georges}},
  \bibinfo{author}{\bibfnamefont{G.}~\bibnamefont{Kotliar}},
  \bibinfo{author}{\bibfnamefont{W.}~\bibnamefont{Krauth}}, \bibnamefont{and}
  \bibinfo{author}{\bibfnamefont{M.~J.} \bibnamefont{Rozenberg}},
  \bibinfo{journal}{Rev. Mod. Phys.} \textbf{\bibinfo{volume}{68}},
  \bibinfo{pages}{13} (\bibinfo{year}{1996}).

\bibitem[{\citenamefont{Metzner and Vollhardt}(1989)}]{metzner89}
\bibinfo{author}{\bibfnamefont{W.}~\bibnamefont{Metzner}} \bibnamefont{and}
  \bibinfo{author}{\bibfnamefont{D.}~\bibnamefont{Vollhardt}},
  \bibinfo{journal}{Phys. Rev. Lett.} \textbf{\bibinfo{volume}{62}},
  \bibinfo{pages}{324} (\bibinfo{year}{1989}).

\bibitem[{\citenamefont{Werner et~al.}(2006)\citenamefont{Werner, Comanac, de'
  Medici, Troyer, and Millis}}]{werner06}
\bibinfo{author}{\bibfnamefont{P.}~\bibnamefont{Werner}},
  \bibinfo{author}{\bibfnamefont{A.}~\bibnamefont{Comanac}},
  \bibinfo{author}{\bibfnamefont{L.}~\bibnamefont{de' Medici}},
  \bibinfo{author}{\bibfnamefont{M.}~\bibnamefont{Troyer}}, \bibnamefont{and}
  \bibinfo{author}{\bibfnamefont{A.~J.} \bibnamefont{Millis}},
  \bibinfo{journal}{Phys. Rev. Lett.} \textbf{\bibinfo{volume}{97}},
  \bibinfo{pages}{076405} (\bibinfo{year}{2006}).

\bibitem[{\citenamefont{Gull et~al.}(2011)\citenamefont{Gull, Millis,
  Lichtenstein, Rubtsov, Troyer, and Werner}}]{ctqmc}
\bibinfo{author}{\bibfnamefont{E.}~\bibnamefont{Gull}},
  \bibinfo{author}{\bibfnamefont{A.~J.} \bibnamefont{Millis}},
  \bibinfo{author}{\bibfnamefont{A.~I.} \bibnamefont{Lichtenstein}},
  \bibinfo{author}{\bibfnamefont{A.~N.} \bibnamefont{Rubtsov}},
  \bibinfo{author}{\bibfnamefont{M.}~\bibnamefont{Troyer}}, \bibnamefont{and}
  \bibinfo{author}{\bibfnamefont{P.}~\bibnamefont{Werner}},
  \bibinfo{journal}{Rev. Mod. Phys.} \textbf{\bibinfo{volume}{83}},
  \bibinfo{pages}{349} (\bibinfo{year}{2011}).

\bibitem[{\citenamefont{Gubernatis et~al.}(1991)\citenamefont{Gubernatis,
  Jarrell, Silver, and Sivia}}]{gubernatis91}
\bibinfo{author}{\bibfnamefont{J.~E.} \bibnamefont{Gubernatis}},
  \bibinfo{author}{\bibfnamefont{M.}~\bibnamefont{Jarrell}},
  \bibinfo{author}{\bibfnamefont{R.~N.} \bibnamefont{Silver}},
  \bibnamefont{and} \bibinfo{author}{\bibfnamefont{D.~S.} \bibnamefont{Sivia}},
  \bibinfo{journal}{Phys. Rev. B} \textbf{\bibinfo{volume}{44}},
  \bibinfo{pages}{6011} (\bibinfo{year}{1991}).

\bibitem[{not({\natexlab{a}})}]{note1}
\bibinfo{note}{The numerical value at the lowest studied temperature of 145~K
  is 0.987}.

\bibitem[{not({\natexlab{b}})}]{note3}
\bibinfo{note}{$\boldsymbol{\phi}\cdot\boldsymbol{\phi}=\phi_x\phi_x+\phi_y\phi_y+\phi_z\phi_z$}.

\bibitem[{not({\natexlab{c}})}]{note2}
\bibinfo{note}{The actual strong coupling expression gives always positive
  $K_0$}.

\end{thebibliography}
\end{document}